\documentstyle{article}

\renewcommand{\arraystretch}{1.5}

\newcommand{\beq}{\begin{equation}}
\newcommand{\eeq}{\end{equation}}
\newcommand{\bea}{\begin{eqnarray}}
\newcommand{\eea}{\end{eqnarray}}
\newcommand{\bi}{\bibitem}
\newcommand{\p}{\partial}

\begin{document}

\begin{center}
              {\LARGE Shear-free, Irrotational, Geodesic, \\
         	      Anisotropic Fluid Cosmologies}
\end{center}

\begin{center}{\Large
		     Des J. Mc Manus
                     and   Alan A. Coley}
\end{center}

\vspace{1.4cm}

\begin{center}{\sl
                 Department of Mathematics, Statistics and Computing Science,\\
                 Dalhousie University,
                 Halifax, NS, Canada B3H 3J5
              }
\end{center}

\vspace{.4cm}

%                         ABSTRACT

\begin{center}{\large {\bf Abstract}} \end{center}

\noindent
General relativistic anisotropic fluid models whose fluid flow lines
form a shear-free, irrotational, geodesic timelike congruence are examined.
These models are of Petrov type D, and are assumed to
have zero heat flux and an anisotropic stress tensor that
possesses two distinct non-zero eigenvalues. Some general results concerning
the form of the metric and the stress-tensor for these models are established.
Furthermore, if the energy density and the isotropic pressure, as measured by
a comoving observer,
satisfy an equation of state of the form
$p = p(\mu)$, with $\frac{dp}{d\mu} \neq -\frac{1}{3}$, then
these spacetimes  admit a foliation by spacelike hypersurfaces
of constant Ricci scalar. In addition, models for
which both the energy density and the anisotropic pressures only depend
on time  are investigated; both spatially homogeneous and spatially
inhomogeneous models are found. A classification of these models
is undertaken. Also,
a particular class of anisotropic fluid models which are  simple
generalizations of the homogeneous isotropic cosmological models is studied.

\vspace{.8cm}

\begin{center} PACS: 04.20.Jb; 98.80.Hw \end{center}

\vspace{.8cm}
\noindent To appear in  {\it Classical and Quantum Gravity 1994}

%\noindent Short Title: Shear-free, Irrotational, Geodesic Fluid Cosmologies

\pagebreak

%%%%%%%%%%%%%%%%%%%%%%%%%%%%%%%%%%%%%%%%%%%%%%%%%%%%%%%%%%%%%%%%%%%%%%%%%%%%%%%
%                                                                             %
                          \section{Introduction}                              %
%                                                                             %
%%%%%%%%%%%%%%%%%%%%%%%%%%%%%%%%%%%%%%%%%%%%%%%%%%%%%%%%%%%%%%%%%%%%%%%%%%%%%%%

In a recent paper \cite{COLEY&MCMANUS}, a
 comprehensive analysis of general relativistic
spacetimes which admit a shear-free, irrotational
and geodesic (SIG) timelike congruence was undertaken. This paper concentrates
on a particular class of such spacetimes, namely,
anisotropic fluid models whose fluid flow lines form a SIG timelike
congruence (SIGA models).
SIGA spacetimes are characterised by a single scale factor and a purely
spatial three-metric  \cite{COLEY&MCMANUS}.
If the additional assumption is made that the energy density and the isotropic
pressure, as measured by a comoving observer,
 satisfy an equation of state of the form $p = p(\mu)$, with $\frac{dp}{d\mu}
\neq -\frac{1}{3}$, then the Ricci scalar associated with the spatial
three-metric is constant; we note that these spacetimes are
not the same as the perfect fluid Friedman-Robertson-Walker (FRW) models,
where the three-space is a space of
constant curvature. However, the FRW models are a subclass of the
SIGA models. The SIGA models studied in this paper
are assumed to have zero heat flux and a non-zero
anisotropic stress tensor, $\pi_{ab}$, as measured by a comoving observer.
The usual phenomenological assumption
$\pi_{ab} = - \lambda \sigma_{ab}$ does not hold for these models since the
fluid congruence has zero shear.
However, there are a variety of physical situations that are described
by such  SIGA models, for example, the interaction of
a perfect fluid and a pure magnetic field
\cite{COLEY&MCMANUS,CHEW,TENERIO&HAKIM,MAARTENS&MASON}.
Furthermore, since the anisotropic stress tensor
is trace-free and three-dimensional in nature,
it must either have three or two distinct non-zero
eigenvalues.  When there are only two distinct eigenvalues then
the matter content of a SIGA spacetime is
that of an anisotropic fluid;  this paper is devoted to the study of such
models.

The paper is organised as follows: In section 2, we derive some general
results about the form of the anisotropic stress tensor. The anisotropic
stress tensor is shown to only depend on a single function, $P$, of the
spatial coordinates, a three-dimensional vector that has unit norm with
respect to the three-dimensional spatial metric and the spatial three-metric.
The spacetime is also
determined to be of Petrov type D.
In sections 3 and 4, we specialise to the case where the three-dimensional
hypersurfaces have constant Ricci scalar, ${}^3R$.
In section 3, we use a 2+1 split of the three-geometry to investigate
a particular class of anisotropic fluid solutions which are  simple
generalizations of the FRW models.
In section 4, we consider the class of solutions
that have constant $P$ and ${}^3R$,
and prove that both spatially homogeneous and spatially inhomogeneous
solutions exist. We finish
by deriving some specific spatially homogeneous solutions for the case where
both  $P$ and ${}^3R$ are constant.

%%%%%%%%%%%%%%%%%%%%%%%%%%%%%%%%%%%%%%%%%%%%%%%%%%%%%%%%%%%%%%%%%%%%%%%%%%%%%%%
%                                                                             %
           \section{The Stress-Energy Tensor and the Metric}                  %
%                                                                             %
%%%%%%%%%%%%%%%%%%%%%%%%%%%%%%%%%%%%%%%%%%%%%%%%%%%%%%%%%%%%%%%%%%%%%%%%%%%%%%%

We consider general relativistic spacetimes with
the following anisotropic fluid stress-energy
tensor\footnote{We shall follow the notation
and conventions in Ellis \cite{ELLIS} and in \cite{COLEY&MCMANUS};
in particular, all kinematical quantities
are defined therein.  Also, Latin indices range
from $0$ to $3$ and Greek indices from $1$ to $3$, and
subscripts indicate differentiation with
respect to the
relevant spacetime coordinates.}
:
\beq
    T_{ab} \;=\; \mu u_a u_b \;+\; p_{\parallel} n_a n_b
		 \;+\; p_{\perp}(u_a  u_b \;-\; n_a n_b \;+\; g_{ab})
    \;\;\;, \label{1}
\eeq
where $u^a$ is a unit timelike vector and  $n^a$ is a unit spacelike
vector orthogonal to $u^a$, $u_a n^a = 0$.
The scalars $p_\parallel$ and $p_\perp$ are the pressures parallel and
perpendicular to $n^a$, respectively, and $\mu$ is the energy-density
as measured by an observer moving with four-velocity $u^a$.
Decomposing   (\ref{1}) with respect to
 $u^a$\cite{COLEY&MCMANUS,ELLIS,ELLIS_1},
we find that the energy flux relative to $u^a$ is zero,
and that  the isotropic pressure and anisotropic stress tensor (as
measured by $u^a$) are given by
\bea
    p        & = & \frac{1}{3} (p_{\parallel} \,+\, 2 p_{\perp}) \label{3}
				 \;\;\;, \\
    \pi_{ab} & = & (p_{\parallel} - p_{\perp}) \{ n_a n_b \,-\,
		\frac{1}{3} (g_{ab} \,+\, u_a\, u_b)\} \;\;\;. \label{4}
\eea

In particular, when the timelike congruence formed by $u_a$ is
geodesic, shear-free, and twist-free then the line
element of the spacetime may be written as \cite{COLEY&MCMANUS}
\beq
     ds^2 \;=\; - dt^2 \,+\, H^2(t) \, h_{\alpha\beta}(x^{\gamma}) dx^\alpha
		  dx^\beta \label{5} \;\;\;.
\eeq
The anisotropic stress-tensor (\ref{4}) is now independent of $t$
and is defined entirely in terms of the Ricci tensor, ${}^3R_{\alpha\beta}$,
associated with the three-metric $h_{\alpha\beta}$:
\beq
    \pi_{\alpha\beta} \;=\;  {}^3R_{\alpha\beta}     \;-\; \frac{1}{3}
			      h_{\alpha\beta} \,{}^3R \label{6} \;\;\;.
\eeq
Furthermore, $\pi_{\alpha\beta}$ must satisfy the conservation equation
\beq
     \nabla^\alpha \pi_{\alpha\beta} \;=\; -\frac{1}{3}   \nabla_\beta
      \,{}^3R  \label{7} \;\;\;,
\eeq
where
$\nabla_\alpha$ is the covariant derivative with respect to the three-metric
$h_{\alpha\beta}$.

In terms of the coordinates (\ref{5}) the vector $n_a = (0, n_\alpha)$
where $n_\alpha$ satisfies
$h^{\alpha\beta} n_\alpha n_\beta = H^{2}(t)$. Thus, the only non-zero
components of the anisotropic stress tensor are
\beq
     \pi_{\alpha\beta} \;=\; (p_\parallel - p_\perp) \{n_\alpha n_\beta
     \,-\, \frac{1}{3} H^2(t) h_{\alpha\beta}(x^\gamma)\} \label{8} \;\;\;.
\eeq
The energy density $\mu$ and the isotropic
pressure $p$ are both functions of $t$
only \cite{COLEY&MCMANUS}, and are related to $H$  by
\bea
     \mu  & = & \frac{1}{3}\, \theta^2 \;+\; \frac{{}^3R}{2 H^2}
	  \;=\;  \frac{6 \dot{H}^2 \,+\, {}^3R}{2 H^2} \;\;\;, \label{MU} \\
     p    & = & - \frac{\dot{\mu}}{\theta} \;-\; \mu
	  \;=\;  - \frac{2 \ddot{H}}{H} \;-\; \frac{6 \dot{H}^2 + {}^3R}{6 H^2}
     \;\;\;, \label{MEAN_P}
\eea
where $\theta(t) = 3 \dot{H}/H$ is
the expansion of the timelike congruence.

\noindent {\bf Theorem 1:}
If the stress-energy tensor has the form (\ref{1}) where the integral curves
of $u^a$ form a shear-free, irrotational, geodesic timelike congruence
then there exists a coordinate system in which the metric has the form
(\ref{5}). In addition, the non-zero components of the anisotropic stress
tensor (in these coordinates) are given by (\ref{8})  where
\bea
    (i) & &
          p_\parallel - p_\perp \;=\; P(x^\alpha)/H^2(t) \;\;\;, \\
    (ii) & &
	      n_\alpha = H(t) N_\alpha(x^\beta) \;\;\;,
\eea
such that  $N_\alpha$ is a unit vector with respect to the spatial
three-metric $h_{\alpha\beta}$.

\noindent {\bf Proof:} Without loss of generality the three-metric can be
diagonalised \cite{PETROV,COTTON}. We take the metric $h_{\alpha\beta}$ to
have the form
\beq
    h_{\alpha\beta} \;=\; {\rm diag} (A^2, B^2,
				      C^2) \;\;\;, \label{9}
\eeq
where $A, B, C$ are independent of $t$.
At least one component $n_\alpha$ is non-zero, say $n_1 \neq 0$.
If $n_2 = 0$ (and$/$or $n_3 = 0$) then the component $\pi_{22}$ implies
$(p_\parallel - p_\perp) H^2 = P(x^\gamma)$ where $P$ is as yet some
undetermined function.
If $n_2 \neq 0$ and $n_3 \neq 0$ then $\pi_{\alpha\beta}(x^\gamma) =
(p_\parallel - p_\perp) n_\alpha n_\beta$ when $\alpha \neq \beta$.
Therefore, there exist functions $f_\alpha{}^\beta(x^\gamma)$ such that
$n_\alpha/n_\beta = f_\alpha{}^\beta$. Thus, the off-diagonal components
of $\pi_{\alpha\beta}$ may be written as
\beq
     \pi_{\alpha\beta} \;=\;
       (p_\parallel - p_\perp) (n_\alpha)^2 f_\beta{}^\alpha
             \;\;\;\;\;\; (\alpha \neq \beta) \;\;\;.
\eeq
(There is no summation over the repeated index $\alpha$ in the above equation.)
Hence,  there exists non-zero functions $G_\alpha(x^\gamma)$
such that
\beq
    (p_\parallel - p_\perp) (n_\alpha)^2 \;=\; G_\alpha(x^\gamma)
    \;\;\;. \label{10}
\eeq

Two possibilities now exist. Either $\pi_{11} \neq 0$ or $\pi_{11} = 0$.
In both cases, (\ref{8}) together with (\ref{9}) and (\ref{10}) implies
there exists a function $P = P(x^\gamma)$ such that
$(p_\parallel - p_\perp) H^2 = P$.

Part $(ii)$ of the theorem now follows directly from part $(i)$ and
expression (\ref{8}) for $\pi_{\alpha\beta}$. \hfill $\Box$

\noindent{\bf Corollary 1:} The anisotropic  stress tensor may be written as
\beq
    \pi_{\alpha\beta} \;=\; P(x^\gamma) \{N_\alpha N_\beta \,-\,
	\frac{1}{3} h_{\alpha\beta}\} \;\;\;, \label{11}
\eeq
where $N_\alpha$ and $P$ satisfy $h^{\alpha\beta} N_\alpha N_\beta = 1$,
and
\beq
     \frac{1}{6} \nabla_\alpha ({}^3R) \;=\;
     \nabla_\beta(P N^\beta) N_\alpha
     \,-\, \frac{1}{3} \nabla_\alpha P\,+\,
     P N^\beta \nabla_\beta N_\alpha
     \label{Cons_P} \qquad \qquad \Box
\eeq

The anisotropic stress tensor $\pi_{\alpha\beta}$ has only two distinct
eigenvalues which implies that the Weyl tensor for metric (\ref{5}) is
Petrov type D \cite{TRUMPER,FMP}.  Conversely, if the metric is of the form
(\ref{5})   and $\pi_{\alpha\beta}$ has two distinct
eigenvalues then $\pi_{\alpha\beta}$ must necessarily be of the
form (\ref{11}), and (\ref{5}) can be interpreted as an anisotropic fluid,
with four-velocity $u^a = \delta^a_t$, whose flow lines are
shear-free, twist-free, and geodesic.
The proof  is essentially as follows: let $e_{(A)}^\alpha$ be the two linearly
independent eigenvectors corresponding to the same eigenvalue, $\lambda$. Then
the  eigenvalue of the third eigenvector, $e_{(1)}^\alpha$, is $-2 \lambda$
since $\pi_{\alpha\beta}$ is traceless. Furthermore, $e_{(1)}^\alpha$ is
orthogonal to $e_{(A)}^\alpha$. If $e_{(1)}^\alpha$ is chosen to have unit norm
then $\pi_{\alpha\beta}$ can be decomposed as
\beq
    \pi^{\alpha\beta} \;=\; \lambda\, \{-2 \,e_{(1)}^\alpha \, e_{(1)}^\beta
                   \,+\, g^{AB}\, e_{(A)}^\alpha \, e_{(B)}^\beta\} \;\;\;,
\eeq
where $g^{AB} \, g_{BC} = \delta^A_C$ and $g_{AB} = e_{(A)} \cdot e_{(B)}$.
The desired result follows from the fact that the three-metric,
$h_{\alpha\beta}$,  has the decomposition
\beq
    h_{\alpha\beta} \;=\;  e_{(1)\alpha} \, e_{(1)\beta}
                   \,+\, g^{AB}\, e_{(A)\alpha} \, e_{(B)\beta} \;\;\;.
\eeq

\setcounter{equation}{0} % Reset the equation counter

%%%%%%%%%%%%%%%%%%%%%%%%%%%%%%%%%%%%%%%%%%%%%%%%%%%%%%%%%%%%%%%%%%%%%%%%%%%%%%%
%                                                                             %
                   \section{A Particular Class of Solutions}
%
%                                                                             %
%%%%%%%%%%%%%%%%%%%%%%%%%%%%%%%%%%%%%%%%%%%%%%%%%%%%%%%%%%%%%%%%%%%%%%%%%%%%%%%

We wish to solve the three-dimensional field
equations (\ref{6}) and (\ref{11}).
As a general prescription, we write the field equations in terms of
a 2+1 split. We suppose that we can choose coordinates $(x, x^A)$
such that $N_\beta \propto \p_\beta x$. (This will always be possible if
$N_\alpha$ is hypersurface orthogonal.) We define base vectors
$e^\alpha_{(A)}= \p_A x^\alpha$ and a lapse function $M$ and
a shift vector $M^A$ via
\beq
    \frac{\p x^\alpha}{\p x} \;=\; M N^\alpha \,+\, M^A e^\alpha_{(A)}
     \;\;\;.
\eeq
The extrinsic curvature and the intrinsic metric for the surface $x =$const
are $K_{AB}$ and $g_{AB}$, respectively. They are related by
\beq
    \p_x g_{AB} \;=\; 2M K_{AB} \,+\, 2 \nabla_{(A} M_{B)} \;\;\;.
    \label{K_AB}
\eeq
It is unlikely that a general solution can be found.
Hence, we  impose some further
physical constraints so that the matter content of the spacetimes under
consideration is
physically relevant. Specifically, we consider the case were
 the energy density and the isotropic pressure satisfy an equation
of state of the form $p = p(\mu)$. Thus, either (i) ${}^3R =$const or (ii)
$\frac{dp}{d\mu} = -\frac{1}{3}$ (see reference \cite{COLEY&MCMANUS}).
We shall only consider the former case here, ${}^3R = R_0 =$ const.
(For a discussion of the latter case see \cite{COLEY&MCMANUS}.)
In addition, we shall only  look for solutions that have unit lapse and
zero shift.

Using the field equations (\ref{6}) and (\ref{11}), and decomposing
the Ricci tensor in terms of a 2+1 split \cite{ADM,ADM_1}, we obtain the
following equations:
 \bea
     {}^3R_{AB} \;\equiv & {}^2R_{AB} \,+\, 2 K_A{}^C K_{BC} \,-\,
                      K K_{AB}    \,-\, \p_x K_{AB}
                 & =\;  (\,{}^3R \,-\, P) g_{AB}/3
		    \label{R1} \;\;\;,\\
     {}^3R_{AN} \;\equiv &\nabla_B K_A{}^B \,-\, \p_A K  &=\; 0
                    \label{R2} \;\;\;,\\
     {}^3R_{NN} \;\equiv &  - K_{AB} K^{AB}  \,-\, \p_x K   &=\;
                     (2 \,P \,+\,  \,{}^3R)/3
                     \label{R3}\;\;\;,
\eea
where $K$ is the trace of the extrinsic curvature.
The Ricci scalar is given by
\beq
    {}^3R \;=\; {}^2R \,-\, K^2 \,-\, K_{AB} K^{AB} \,-\, 2 \,
		 \p_x K
     \;\;\;. \label{Ricci}
\eeq
(Note that the Ricci tensor for any two metric is
${}^2R_{AB} =   {}^2R\, g_{AB}/2$.)

Under the conditions we have assumed here,
equation (\ref{K_AB}) can be integrated to yield
\beq
    g_{AB} \;=\; \Phi(x, x^A) \, h_{AB}(x^C)  \label{g_AB}
\eeq
with $2\lambda = \p_x \Phi /\Phi$.
Equation (\ref{R2}) implies that $\lambda = \lambda(x)$ which
in turn yields $\Phi = \psi(x) \, \phi(x^a)$. We can absorb $\phi$
into $h_{AB}$. Thus, without loss of generality, the  two-metric (\ref{g_AB})
can be expressed as
\beq
    g_{AB} \;=\; \psi(x)\, \phi(x^C)\, \delta_{AB} \label{g1}\;\;\;.
\eeq
Now the function $2\lambda = \psi^\prime/\psi$.
(We have introduced the notation
$\psi^\prime \equiv \p_x \psi$.) For the remainder of the paper, we shall
take $x^A = (y, z)$.

The equation for the Ricci scalar (\ref{Ricci})  reduces to
\beq
    R_0 \;=\; {}^2R \,-\, 6 \lambda^2  \;-\; 4  \lambda^\prime \;\;\;.
    \label{R_0}
\eeq
The Ricci scalar for the metric (\ref{g1}) is easily calculated since
the metric is conformally flat \cite{WALD}, and is given by
\beq
    {}^2R \;=\;  -\frac{\nabla^2 \, {\rm ln} \phi}{\psi \phi}
    \label{2R}
\eeq
($\nabla^2$ is simply the flat space Laplacian).
But (\ref{R_0}) implies that ${}^2R = {}^2R(x)$. Therefore, $\phi$
must satisfy
\beq
    \nabla^2 {\rm ln} \phi \,+\, k \phi \;=\; 0 \label{phi_k}
\eeq
where $k$ is some constant. Solving (\ref{phi_k}) is equivalent to finding
the conformal factor for a two-space of constant curvature.
A rescaling of the  $y$ and $z$ coordinates enables us to write
the general
solution as \cite{EISENHART2}
\beq
    \phi \;=\; [1 \,+\, \frac{\kappa}{4} (y^2 \,+\, z^2)]^{-2} \label{phi}
    \;\;\;,
\eeq
where the constant $\kappa$ has been normalised to take values $\pm 1$ and $0$.

Thus, inserting the results (\ref{phi}) and (\ref{2R}) into (\ref{R_0}),
we obtain the following differential equation for $\psi$:
\beq
    \psi \, \psi^{\prime\prime} \,-\, \frac{1}{4} (\psi^\prime)^2
    \,+\, \frac{R_0}{2} \psi^2 \;=\; \kappa \psi \;\;\;.
\eeq
We introduce a new function $S(x)$
by letting $\psi = S^2$.
The function $S$ then satisfies the differential
equation
\beq
    2 S S^{\prime\prime} \,+\, (S^\prime)^2 \,+\, \frac{R_0}{2} S^2
    \;=\; \kappa \;\;\;. \label{RR_eqn}
\eeq
The first integral of the above equation is
\beq
    (S^\prime)^2 \;=\; \kappa \,+\, \frac{2\alpha}{S} \,-\,
		       \frac{R_0}{6} S^2 \label{R_eqn}\;\;\;,
\eeq
where $\alpha$ is an integration constant.
Equation (\ref{R_eqn})
is very reminiscent of the Friedmann equation. Indeed an equation
similar to (\ref{R_eqn}) has been studied in the context of
homogeneous and isotropic
cosmological models \cite{ROBERTSON,HARRISON}.

An expression for $P$ is found by subtracting (\ref{R1}) from (\ref{R3}):
\beq
    P \;=\; - \frac{S^{\prime\prime}}{S} \,+\,
    \left( \frac{S^\prime}{S} \right)^2 \,-\, \frac{\kappa}{S^2}
    \;\;\;. \label{P_1}
\eeq
Equations (\ref{RR_eqn}) and (\ref{R_eqn}) then imply
\beq
     P \;=\; \frac{3 \alpha}{S^3} \;\;\;,
\eeq
which is in complete agreement with the conservation equation (\ref{Cons_P}).

Thus, the metric
\beq
    ds^2 \;=\; - dt^2 \,+\, H^2(t) \left\{ dx^2 \,+\, S^2(x)
    \frac{dy^2 \,+\, dz^2}{[1 \,+\, \frac{\kappa}{4}(y^2 + z^2)]^2}
    \right\}
\eeq
can be interpreted as  an anisotropic fluid, with four-velocity
$u^a = \delta^a_t$, when $\kappa = \pm 1$ or $\kappa = 0$,
and $S(x)$ satisfies (\ref{R_eqn}).
The energy density, $\mu$ and the isotropic pressure, $p$,
as measured by a comoving observer are given by (\ref{MU}) and
(\ref{MEAN_P}), respectively.
The pressures $p_\perp$ and $p_\parallel$ are
\bea
    p_\perp  &=& p \,-\, \frac{\alpha}{H^2 \, S^3} \;\;\;, \\
    p_\parallel      &=& p \,+\, \frac{2\alpha}{H^2 \, S^3} \;\;\;.
\eea

We note that if $\alpha$ equals zero then the model does not represent an
anisotropic fluid model since  $P = 0$. This implies that
$p_\parallel = p_\perp$, and thus, the stress-energy tensor (\ref{1})
corresponds to a perfect fluid and the spacetime is consequently
a perfect fluid FRW model. Therefore, we shall only consider the case of
non-zero $\alpha$.

A close examination of (\ref{R_eqn}) reveals that it is remarkably similar
to the zero pressure Friedmann equation. However, there is one
notable exception; the constant $\alpha$ must be strictly non-negative
for the analogue
to be complete since $\alpha$ basically corresponds to the energy density,
$\rho$, in the Friedmann equation. In our analysis, $\alpha$ simply
represents an arbitrary integration constant, and thus may take on
negative values.

Equation (\ref{R_eqn}) is difficult to analyse in general. However,
a qualitative analysis can easily be undertaken.
First, we shall consider solutions of (\ref{R_eqn}) that allow $S(x)$ to
tend to zero as unphysical.
In other words, there exists a number
$S_{\rm min}$ such that $0 < S_{\rm min} \leq S(x)$ for all values of $x$.
In addition, if $R_0 > 0$ then $S(x)$ must also be bounded from above,
that is, there exists a number $S_{\rm max} < \infty$ such that
$S(x) \leq S_{\rm max}$.

In terms of the Harrison-Robertson classification
scheme \cite{ROBERTSON,HARRISON},
there are four basic types of solutions for which $S(x)$ is non-zero
everywhere.
These solutions are referred to as:
$(i)$ static $S_1$ (unstable) and  $S_2$ (stable),  where $S(x)=S_s=$const and
$S^\prime = S^{\prime\prime} = 0$;
$(ii)$ asymptotic $A_2$, $S_s \leq S \leq \infty$ where $S_s$ is a static
point,
that is, $S^\prime(S = S_s)  = S^{\prime\prime}(S = S_s) = 0$;
$(iii)$ monotonic $M_2$,
$S_{\rm min} \leq S \leq \infty$ with $S(x = \pm\infty) = \infty$; and
$(iv)$ oscillatory $O_2$, where $S_{\rm min} \leq S \leq S_{\rm max}$.
The allowable solutions are summarised in Table 1.

The static models are by far the simplest models.
Equations (\ref{RR_eqn}) and (\ref{R_eqn}) imply that
$S^2 = 2 \kappa/R_0$ (provided $R_0 \neq 0$ and $\kappa \neq 0$) and
$\alpha = \kappa R_0/3$. Thus, sign$(R_0) = \kappa$.
Equation (\ref{P_1}) then gives
$P = -R_0/2 =$const.
Thus, the metric for the  static models ($\kappa = \pm1$) is
\beq
    ds^2 \;=\; - dt^2 \,+\, H^2(t) \left\{ dx^2 \,+\,
    \frac{dy^2 \,+\, dz^2}{[1 \,+\, \frac{\kappa}{4}(y^2 + z^2)]^2}
    \right\} \;\;\;. \label{ds_STATIC}
\eeq
(The coordinates have been rescaled to normalise the Ricci scalar, $R_0$,
to $2\kappa$.)
The above metric is a Kantowski-Sachs metric \cite{K&S} with equal scale
factors.
The energy density and the isotropic
pressure are given by (\ref{MU}) and (\ref{MEAN_P}), respectively, viz.
\bea
     \mu  & = &
	    \frac{3 \dot{H}^2 \,+\, \kappa}{ H^2} \;\;\;,  \\
     p    & = &
	  - \frac{6 H \ddot{H} \;+\; 3 \dot{H}^2 + \kappa}{3 H^2}
     \;\;\;.
\eea
The anisotropic pressures associated with the  metric (\ref{ds_STATIC}) are
\bea
    p_\parallel  &=&  -\, \frac{2 \ddot{H} H + \dot{H}^2}{H^2} \;\;\;, \\
    p_\perp      &=& -\, \frac{2 H \ddot{H} +  \dot{H}^2 + \kappa}{ H^2}
    \;\;\;.
\eea

 The following anisotropic fluid with planar symmetry is another particular
example:
\beq
    ds^2 \;=\; -dt^2 \,+\, H^2(t) \{dx^2 \,+\, S^2(x)(dy^2 \,+\, dz^2)\}
\eeq
and corresponds to taking $\kappa = 0$. The only non-singular solution
occurs when $R_0 < 0$ and $\alpha < 0$, and then
\beq
    S(x) \;=\; \left(\frac{12 \alpha}{R_0}\right)^{\frac{1}{3}}
	       \left[ \cosh\left\{\sqrt{\frac{-3 R_0}{8}} x\right\}
	       \right]^{\frac{2}{3}}
	       \;\;\;.
\eeq

\setcounter{equation}{0} % Reset the equation counter

%%%%%%%%%%%%%%%%%%%%%%%%%%%%%%%%%%%%%%%%%%%%%%%%%%%%%%%%%%%%%%%%%%%%%%%%%%%%%%%
%                                                                             %
                \section{Solutions with constant $P$}
 %
%                                                                             %
%%%%%%%%%%%%%%%%%%%%%%%%%%%%%%%%%%%%%%%%%%%%%%%%%%%%%%%%%%%%%%%%%%%%%%%%%%%%%%%

In this section, we shall look for solutions that have both ${}^3R$ and $P$
constant.  First, we note  that  this assumption greatly simplifies the
conservation equations (\ref{Cons_P}). However,  if $P$ is not constant
then the resulting models are inhomogeneous; consequently a more pertinent
reason for choosing this ansatz
is the fact that we would like to determine the spatially homogeneous
SIGA models.

We shall assume  that the matter associated with a spatially
homogeneous model inherits its symmetries \cite{COLEY&TUPPER};
in particular,
any scalars  constructed from the stress-energy tensor
should be independent of the spatial coordinates.
For an  anisotropic fluid this would imply that
$P$  is constant since $P = P(x^\alpha)$.
The vector $N_\alpha$ must be both geodesic,
$N^\beta \nabla_\beta N_\alpha =0$,
and expansion-free, $ \nabla^\alpha N_\alpha = 0$,
with respect to the three-metric $h_{\alpha\beta}$ (if $P =$ const and
${}^3R = R_0$).
We already know that there exist spatially homogeneous SIGA
models with constant $P$ and ${}^3R$ (see \cite{M&C}).
Thus, we now ask the following two questions:(1) are the Mimoso and Crawford
solutions \cite{M&C} the only spatially homogeneous SIGA models, and (2) are
there any spatially inhomogeneous models with
$P$ and ${}^3R$ constant?

Thus, suppose that  the Ricci tensor has the form
\beq
    {}^3R_{\alpha\beta} = (\alpha - \beta) \, N_\alpha \, N_\beta \;+\;
     \beta \, h_{\alpha\beta} \;\;\;, \label{ds3}
\eeq
where the eigenvalues, $\alpha$ and $\beta$, are constant and
$N_\alpha$ is a  unit vector.
The contracted Bianchi identities imply that $N_\alpha$ can only have
twist and shear, and hence
\beq
     \nabla_\beta N_\alpha \;=\; \omega_{\alpha\beta} \,+\,
                              \sigma_{\alpha\beta} \;\;\;,
\eeq
where  $\omega_{\alpha\beta}  = \nabla_{[\beta} N_{\alpha]}$
is the twist tensor and $\sigma_{\alpha\beta}  = \nabla_{(\beta} N_{\alpha)}$
is the shear tensor.
If the shear tensor is zero then a theorem due to Bona and Coll
\cite{B&C1,B&C2}
implies that the three-metric $h_{\alpha\beta}$
admits a $G_4$ as the maximal isometry group. If the
shear is non-zero then we can use its unit orthogonal eigenvectors to
 form an orthogonal triad
$\{ N_\alpha,\, E_{(1)\alpha}, \,E_{(2)\alpha}\}$.
The shear and twist can then be
decomposed as
\bea
     \sigma_{\alpha\beta} &=& \sigma \, \left[ E_{(1)\alpha} \, E_{(1)\beta}
                    \,-\, E_{(2)\alpha} \, E_{(2)\beta}\right] \;\;\;, \\
      \omega_{\alpha\beta} &=& \omega  \, \left[ E_{(1)\alpha} \, E_{(2)\beta}
                    \,-\, E_{(2)\alpha} \, E_{(1)\beta}\right] \;\;\;,
\eea
where the vorticity and shear scalars are given by
$2 \,\omega^2 = \omega_{\alpha\beta} \, \omega^{\alpha\beta}$ and
$2 \,\sigma^2 = \sigma_{\alpha\beta} \, \sigma^{\alpha\beta} \neq 0$,
respectively. The integrability conditions for $N_\alpha$,
\beq
     \nabla_\gamma \nabla_\beta N_\alpha \,-\,
     \nabla_\beta \nabla_\gamma N_\alpha  \;=\;
     N^\delta \, R_{\delta\alpha\beta\gamma} \label{eq4} \;\;\;,
\eeq
can be employed to show that
\beq
    \omega^\prime \;=\; \sigma^\prime \;=\;
     (E_{(1)\alpha})^\prime \;=\; (E_{(2)\alpha})^\prime \;=\; 0 \;\;\;,
\eeq
where $ (\cdots)^\prime \equiv N^\alpha \nabla_\alpha (\cdots)$ is
the covariant derivative in the direction of $N^\alpha$. Decomposing
the covariant derivatives of $E_{(A)\alpha}$ in terms of the  triad,
we find
\bea
    \nabla_\beta E_{(1)\alpha} &=& - \theta_2 \, E_{(2)\alpha} \, E_{(1)\beta}
       \,+\, \theta_1 \,  E_{(2)\alpha} \, E_{(2)\beta} \,-\,
        \sigma \, N_\alpha \,  E_{(1)\beta} \,-\,
        \omega  \, N_\alpha \,  E_{(2)\beta} \;\;\;, \\
   \nabla_\beta E_{(2)\alpha} &=& - \theta_1 \, E_{(1)\alpha} \, E_{(2)\beta}
       \,+\, \theta_2  \, E_{(1)\alpha} \, E_{(1)\beta} \,+\,
        \omega \,N_\alpha \,  E_{(1)\beta} \,+\,
        \sigma  \, N_\alpha \,  E_{(2)\beta} \;\;\;,
\eea
where $\theta_1$ and $\theta_2$ are the expansions of $E_{(1)}$ and $E_{(2)}$,
respectively. Equation (\ref{eq4}) also  yields three further equations:
\bea
     \alpha \,=\, 2 (\omega^2 \,-\, \sigma^2) & & \;\;\;,  \label{eq9}\\
     \sigma_1 \,-\, \omega_2 \,+\, 2 \, \sigma \, (\theta_1) \,=\, 0 & &
     \;\;\;,
             \label{eq10}  \\
      \sigma_2 \,-\, \omega_2 \,+\, 2 \,\sigma \,(\theta_2) \,=\, 0 & & \;\;\;
             \label{eq11} ,
\eea
where $\sigma_1 =  E^\alpha_{(1)} \, \nabla_\alpha \sigma, \,
\sigma_2 = E^\alpha_{(2)} \, \nabla_\alpha \sigma$ , etc.

Taking
$\{ N_\alpha \, dx^\alpha , E_{(1)\alpha} \, dx^\alpha,
E_{(2)\alpha} \, dx^\alpha\}$
as an orthogonal basis of 1-forms for the metric and using Cartan's
structural equations, we obtain the extra integrability conditions:
\bea
    (\theta_2)^\prime \,+\, (\theta_2) \sigma \,+\, (\theta_1) \omega &=& 0
          \label{eq12}  \;\;\;, \\
    (\theta_1)^\prime \,-\, (\theta_2) \omega \,-\, (\theta_1) \sigma &=& 0
            \label{eq13} \;\;\;, \\
    (\theta_2)_{,2} \,+\, (\theta_1)_{,1} \,+\, (\theta_2)^2 \,+\,
       (\theta_1)^2 \,+\, \beta &=& 0 \;\;\;. \label{eq14}
\eea
We now notice that (\ref{eq12}) implies
$(\theta_2)^{\prime\prime} = -\frac{\alpha}{2} (\theta_2)$.
Introducing coordinates $ x^\alpha = (x, y^A)$  such that  $x$ is the affine
distance along the integral curves of $N^\alpha$, (\ref{eq12}) and (\ref{eq13})
can then be integrated.
At this stage the analysis splits into two parts: (i) $\alpha \neq 0$ and (ii)
$\alpha = 0$.

\noindent {\bf (i)}:
If $\alpha \neq 0$ then (\ref{eq14}) can be employed  to show that the
only valid solution is $\theta_1 = \theta_2 = 0$, whence $\beta = 0$. Equations
(\ref{eq9})--(\ref{eq11}) then imply that both $\omega$ and $\sigma$ are
constants. Hence, the metric must admit a $G_3$ \cite{B&C1,B&C2} since
all the eigenvalues and spin coefficients are constant (see \cite{DES_JMP}
for a complete classification of all postive definite three-metrics that
satisfy (\ref{ds3}) and admit either a $G_3$ or a $G_4$ as the maximal
isometry  group). Thus, we have proved the following result:

\noindent{\bf Theorem 2:}
If $h_{\alpha\beta}$ is a positive definite Riemannian three-metric whose Ricci
tensor has exactly two distinct constant eigenvalues then the metric
is homogeneous if either
$(i)$ the shear of the principal Ricci direction
corresponding to the simple eigenvalue is zero  or
$(ii)$ the degenerate eigenvalue is zero. \hfill $\Box$

We note that the Killing vectors of the three-space are also Killing vectors
of the four-dimensional spacetime and hence, the resulting SIGA models
are spatially homogeneous.

\noindent {\bf (ii)}
If $\alpha =0$ then we can always choose $\omega = \sigma$, and consequently
(\ref{eq13}) implies $\theta_2 = - \theta_1 = \theta(y^A)$. Equations
(\ref{eq9})--(\ref{eq11}) yield
\bea
     \theta_{,1} \,-\, \theta_{,2} &=& 2 \, \theta^2 \,+\, \beta \;\;\;, \\
     \sigma_{,1} \,-\, \sigma_{,2} &=& 2 \, \sigma \, \theta \;\;\;.
\eea
We note that $N^\alpha$ is \underline{not} hypersurface orthogonal.
However, we can pick a new orthogonal triad
$\{ N^\alpha,\, \xi_{(+)}^{\alpha}, \,\xi_{(-)}^{\alpha}\}$, where
\beq
     \xi^\alpha_{(\pm)} \;=\;
     \left[ E^\alpha_{(1)} \, \pm \, E^\alpha_{(2)} \right] / \sqrt{2} \;\;\;,
\eeq
such that $\xi^\alpha_{(+)}$ is hypersurface orthogonal (we will label
these surfaces by $z=$ const) and such that  $\xi^\alpha_{(-)}$ is
geodesic. If we let $y$ be the affine distance along the integral curves
of  $\xi^\alpha_{(-)}$ then $(x,y)$ will be intrinsic coordinates for
the surfaces $z=$ const and a 2+1 split can be employed to show that the
three-metric can then be written as
\beq
     {}^3ds^2 \;=\; dx^2 \,+\, dy^2 \,+\, 2 M_1 \, dx \, dz + 2 M_2 \, dy \, dz
           \,+\, [M^2 \,+\, (M_1)^2 \,+\, (M_2)^2] \, dz^2 \;\;\;, \label{eq18}
\eeq
where
\bea
     M_1 \;=\; -y \,+\, g(z) &{\rm and}& M_2 \;=\; x \,+\, h(z) \;\;\;,
\eea
and
\beq
    M \;=\; \left\{
      \begin{array}{lcl}
        A(z) \, e^{\sqrt{-\beta} y} \,+\, B(z) \, e^{-\sqrt{-\beta} y}
           &\;\;\;{\rm if}\;\;\;&   \beta < 0  \\
        A(z) \, \cos(\sqrt{\beta} y) \,+\, B(z) \, \sin(\sqrt{\beta} y)
           &\;\;\;{\rm if}\;\;\;&   \beta > 0
      \end{array}
    \right. \label{eq20}
\eeq
where $A, B, g$ and $h$ are arbitrary functions.

\noindent{\bf Theorem 3:}
If $h_{\alpha\beta}$ is a positive definite inhomogeneous Riemannian
three-metric whose Ricci tensor has exactly two distinct constant eigenvalues
then (1) the simple eigenvalue of the Ricci tensor is zero and (2) there
exist coordinates $(x,y,z)$ such that the inhomogeneous metric is given
by (\ref{eq18})--(\ref{eq20}). \hfill $\Box$

{}From theorems 2 and 3, we have the following corollary:

\noindent{\bf Corollary 2:}
If $h_{\alpha\beta}$ is a positive definite Riemannian
three-metric whose Ricci tensor has exactly two distinct constant eigenvalues
then $h_{\alpha\beta}$  is not homogeneous
if and only if  both the simple
eigenvalue of the Ricci tensor is zero and the shear of the principal Ricci
direction corresponding to the simple eigenvalue is nonzero. \hfill $\Box$

We note that these spatially inhomogeneous models may be of some interest in
the study of
anisotropic fluid cosmologies. We also note that the three-metric,
(\ref{eq18}), possesses at most one Killing vector field (none in general)
 and that there exist examples of these spatially inhomogeneous models
 satisfying all of the relevant energy conditions.

\subsection{Spatially Homogeneous Solutions}
Let us explicitly construct some of the spatially homogeneous
 three-metrics. Now, since
all of the spatially homogeneous cosmological models have at least one
translational Killing vector field, say $\xi = \frac{\p}{\p x}$,
we shall consider the following class of three-metrics as an illustrative
example:
\beq
    {}^3ds^2 \;=\; A^2(y,z) \, dx^2 \;+\; B^2(y,z) (dy^2 \;+\; dz^2)
    \;\;\;. \label{DS_KVF}
\eeq
% which contains all but the Bianchi {\tt II} solutions.

The Ricci tensor components
$R_{xy}$ and $R_{xz}$ for the above metric are identically zero,
which implies that $\pi_{xy}=\pi_{xz}=0$.
Therefore, we must have $P \, N_1 \, N_2 = P \, N_1 \, N_3 = 0$.
There are three possible solutions: ($i$) $N_1 \neq 0$ and $N_2=N_3=0$ with
$P \neq 0$, ($ii$) $N_1 = 0$ and $P\neq 0$, and ($iii$) $P=0$.
We can neglect the third possibility since it would imply that the
stress-energy tensor represents a perfect fluid, as was  noted earlier.

\noindent{\large $(i) \; \underline{N_1 \neq 0; \; N_2 = N_3 = 0}$}

The expansion is $\nabla^\alpha N_\alpha \equiv A^{-2} N_{1,x} = 0$, and
the acceleration equations are $\nabla^\alpha N_1 \equiv 0, \,
\nabla^\alpha N_2  \equiv -N_1^2 A^{-3} A_y = 0, \,
\nabla^\alpha N_3 \equiv -N_1^2 A^{-3} A_z = 0$. Hence, we can take $N_1= A =
1$
since $N_\alpha$ is a unit vector. Thus, the metric belongs to the class of
metrics discussed in section 3. Now, the metric must be the static metric
solution (\ref{ds_STATIC}) since the Ricci tensor has constant
eigenvalues (both $P$ and $R$ constant).

\noindent{\large$(ii) \;  \underline{N_1 = 0}$}

Now, $N_1 = N^1 =0$ implies  that there exists a coordinate transformation
$(y,z) \rightarrow (v,w)$  such that $w$ is the affine-distance
along the integral curves of $N^\alpha$; that is, $N^\alpha = \delta^\alpha_w$.
Thus, since $N^\alpha$ is
geodesic the three-metric (\ref{DS_KVF}) may be written as
\beq
    {}^3ds^2 \;=\; A^2(v,w) \, dx^2 \,+\, \alpha^2(v,w) \, dv^2 \,+\,
		   2\, \beta(v)\, dv\,dw \,+\, dw^2 \;\;\;.
\eeq
The coordinate transformations
$w^\prime = w + \int\!\beta(v) dv\,,\,v^\prime = \int\!\beta(v) dv$ reduce
the above metric to the form
\beq
    ds^2 \;=\; A^2(v,w) dx^2 \,+\, B^2(v,w) dv^2 \,+\, dw^2
\eeq
(the primes have been dropped) and maintains the form of $N^\alpha$,
$N^\alpha = \delta_w^\alpha$.

The expansion of $N_\alpha$ is
$\nabla^\alpha N_\alpha \;\equiv\; A^{-1} A_{w} \;+\; B^{-1} B_w \;=\; 0$,
which yields the equation $\p_w (A B) = 0$.
Also, the field equations imply that
$R_{vw} \equiv  - A^{-1} A_{vw} \,+\, A^{-1} B^{-1} A_v B_w = 0$, which implies
that $\p_w (A_v/B) \;=\; 0$. Hence,  we find
\bea
    A^2 &=& F(v) \;+\; G(w) \;\;\;, \\
    B^2 &=& \frac{g^2(v)}{F(v) \;+\; G(w)} \;\;\;,
\eea
where $F,G$, and $g$ are arbitrary functions (we can set $g = 1$
without loss of generality).  Next, we can reduce the field equation
$R_{ww} \equiv  - B^{-1} B_{ww} \,-\, A^{-1} A_{ww}   =  (2 P + R_0)/3$
to
\beq
    \left( \frac{G_w}{F \;+\; G} \right)^2 \;=\;
    -\, \frac{2}{3} (2 P \,+\, R_0) \;\;\;. \label{eq3}
\eeq
Therefore, we need only examine the two cases: (a) $G = 0$ and
(b) $F = 0$.

\noindent (a) $\underline{G = 0}$

Here, $A = A(v) =  B^{-1}$ and
equation (\ref{eq3}) gives $2 P + R_0 = 0$. Thus, $B$ can be set equal to 1.
The field equations
$R_{xx}  = (R_0 - P) A^2/3$ and  $R_{vv}  = (R_0 - P) B^2/3$
both reduce to the single ordinary differential equation
\beq
    A_{vv} \;=\; - \frac{1}{2} R_0 A \;\;\;. \label{A_eq}
\eeq
We can rescale the $(x, v, w)$ coordinates so that $R_0 = 2 \kappa$.
Then, the solution to (\ref{A_eq}) is
\beq
    A \;=\; \left\{
	     \begin{array}{lcl}
	      a \sin(v)  \,+\,
	      b  \cos(v) &\;\;\;{\rm if}\;\;\;& \kappa = 1 \\
	      a \exp(v)  \,+\,
	      b  \exp(-v) &\;\;\;{\rm if}\;\;\;& \kappa = -1
             \end{array} \right.
    \;\;\;,
\eeq
where $a$ and $b$ are arbitrary constants.
(When  $\kappa = 0$  the metric is the flat FRW model.) If $\kappa = -1$
then there are essentially three distinct choices for $A$
\renewcommand{\arraystretch}{1}
\beq
     A \;=\;  \left\{
    \begin{array}{lc}
        (i)   & e^v \\
        (ii)  & \cosh(v) \\
        (iii) & \sinh(v)
    \end{array}
    \right. \;\;\;. \label{eq30}
\eeq
\renewcommand{\arraystretch}{1.5}
However, in each of the above cases, the two-surface $w=$ constant is a
surface of constant negative curvature. Thus the three metrics are
diffeomorphic. The metric admits a $G_4$ with a
simply-transitive $G_3$
subgroup of Bianchi type {\tt III} and a $G_3$ subgroup of Bianchi type
{\tt VIII} acting on the two-dimensional space $w =$ constant.
If $\kappa = 1$ then  the metric is a Kantowski-Sachs \cite{K&S} metric
\beq
    ds^2 \;=\; -dt^2 \;+\; H^2(t) (\sin^2(v) dx^2 \,+\, dv^2 \,+\, dw^2)
    \;\;\;.
\eeq
Again, the three-metric
is  homogeneous and  admits a $G_4$.

%\newpage
\noindent (b) $\underline{F = 0}$

In this case, $A = A(w) = B^{-1}$. The field equations
reduce to:
\beq
\begin{array}{lcrl}
   A^{-2} R_{xx} &=&  - A^{-1} A_{ww} \;+\; A^{-2} A^2_w &=
                (R_0 - P)/3 \;\;\;, \\
   B^{-2} R_{vv} &=&  A^{-1} A_{ww} \;-\; A^{-2} A^2_w &= (R_0 - P)/3 \;\;\;,\\
   R_{ww} &=& -2 A^{-2} A^2_w &= (2 P + R_0)/3 \;\;\;.
\end{array}
\eeq
We observe that $P = R_0$, and that a solution only exists for $R_0 < 0$.
Without loss of generality, we can take $R_0 = -2$, and then we have the
solution $A = \exp(\pm w)$.    The solution is the Bianchi type {\tt VI}${}_0$
metric
\beq
    ds^2 \;=\; -dt^2 \;+\; H^2(t) (e^{2w} dx^2 \,+\, e^{-2w} dv^2 \,+\, dw^2)
    \;\;\;.
\eeq

Mimoso and Crawford \cite{M&C} discussed  Bianchi types {\tt I, III} and
Kantowski-Sachs
metrics in the context of anisotropic fluid models.
However, these are not the only SIGA models that  can be interpreted as
SIGA models. There are also Bianchi type {\tt II, VI, VIII} and {\tt IX}
examples of SIGA models \cite{DES_JMP}.

\setcounter{equation}{0} % Reset the equation counter

%%%%%%%%%%%%%%%%%%%%%%%%%%%%%%%%%%%%%%%%%%%%%%%%%%%%%%%%%%%%%%%%%%%%%%%%%%%%%%%
%                                                                             %
                         \section{Conclusion}                                 %
%                                                                             %
%%%%%%%%%%%%%%%%%%%%%%%%%%%%%%%%%%%%%%%%%%%%%%%%%%%%%%%%%%%%%%%%%%%%%%%%%%%%%%%

We have demonstrated that there exist coordinates such that the metric
of a SIGA spacetime has the form (\ref{5}), where
${}^3R_{\mu\nu}(h_{\gamma\delta}) = (\alpha - \beta) N_\mu N_\nu
  + \beta \, h_{\mu\nu}$.
In particular, we examined models in which both of the eigenvalues of
${}^3R_{\alpha\beta}$ are constant; In this case,
if either $\beta$ or the shear of $N_\mu$  is zero
then the three-metric, $h_{\mu\nu}$, is homogeneous and hence,
the associated SIGA models are spatially homogeneous. A necessary condition
for the three-metric to be inhomogeneous is $\alpha = 0$. However, this
condition is not sufficient. The static models discussed in section 3 are
examples of spatially homogeneous models with $\alpha = 0$.

In this paper and our recent paper \cite{COLEY&MCMANUS}, we have only been
interested in the study of spacetimes which admit a  SIG {\it timelike}
congruences. However, it is also of interest to study spacetimes which
admit a SIG congruence  which is either {\it null} or {\it spacelike}.
We shall investigate the null case in a future paper. In the spacelike case,
perhaps the most natural (and most simple) setting in which such a
congruence may take on a physical role is that in which the spacelike
congruence is related to the preferred  spacelike direction associated with an
anisotropic
fluid source [see (\ref{1})].  Let us now briefly describe this case since
the analysis involved is similar to that employed in this paper.

Thus, we study anisotropic fluid spacetimes with an energy-momentum tensor
given by (\ref{1}) in which the preferred spacelike vector $n^a$ orthogonal
to $u^a$ is shear-free, irrotational and geodesic. In analogy with
(\ref{5}), there exist coordinates in which $n^a = \delta^a_x$ and the
metric is given by
\beq
    ds^2 \;=\; dx^2 \,+\, L^2(x) \, l_{\xi\zeta}(x^\psi) \, dx^\xi \, dx^\zeta
    \;\;\;, \label{C1}
\eeq
where $\xi, \zeta, \psi  = 0,2,3 \;\; [x^\psi = (t,y,z)]$ and
$l_{\xi\zeta}(x^\psi)$ are the components of a Lorentzian three-metric with
signature $+1$ (and coordinates can always be chosen such that $l_{\xi\zeta}$
is diagonal). Exact solutions of the Einstein field equations can now be
sought in a similar manner to those obtained earlier in this paper,
particularly in the case in which the adapted coordinates (\ref{C1}) are
also comoving coordinates (that is, in the case in which the anisotropic
fluid four-velocity $u^a$ can be aligned along the timelike coordinate axis).
However, we note that these exact solutions will be somewhat different in
form to those obtained earlier due to the Lorentzian signature of
$l_{\xi\zeta}$.
We also note that in the special  perfect fluid subcase these solutions
are related to the solutions of Stephani \cite{STEPHANI}.

In particular, the results of Bona and Coll \cite{B&C3} on
Lorentzian three-metrics can be utilized
in the special case in which the three-dimensional
Ricci tensor, ${}^3R_{\xi\zeta}$, associated with the metric $l_{\xi\zeta}$
has exactly two distinct constant eigenvalues. The results of section 4 can
be generalized  to the Lorentzian case and  hypersurface homogeneous
models can be obtained.

\section*{Acknowledgements:}
This work was supported, in part, by the Natural Science and Engineering
Research Council of Canada.

%                       BIBLIOGRAPHY

\section*{Table 1:}

\begin{tabular}{|c|c|c|l|} \hline
$\kappa$ & $\alpha$ & $R_0$ & Type \\ \hline\hline
0 & $<0$ & $\leq 0$ & $M_2$ \\ \hline
1 & $<0$ & $<0$ & $M_2$ \\ \cline{3-4}
  & $  $ & $>0$ & $O_2, S_2$ \\ \hline
-1& $<0$ & $<0$ & $M_2$ \\ \cline{2-4}
  & $>0$ & $<0$ & $S_1, A_2, M_2$ \\ \hline
\end{tabular}

\vspace{1cm}

 \noindent Table caption: Harrison-Robertson classification
           \cite{ROBERTSON,HARRISON} for anisotropic
           fluid solutions
           with $S(x) \geq S_{\rm min} > 0$. (The definitions of
           the constants $\kappa, \alpha, R_0$ and the spatial types
           can be found in the text.)

\end{document}